\documentclass[journal]{vgtc}                     

\onlineid{1705}

\vgtccategory{Research}

\vgtcpapertype{Data Transformations}

\usepackage{tabu}      
\usepackage{booktabs}

\graphicspath{{./figs/}} 

\usepackage{amsfonts}
\usepackage{amsmath}
\usepackage{multirow}

\newcommand{\para}[1] {\vspace{1pt}\noindent{\textbf{#1}}}

\newcommand{\Mspace}{\mathbb{M}}
\newcommand{\Rspace}{\mathbb{R}}

\newcommand{\update}[1]{\textcolor{black}{{#1}}}
\newcommand{\etal}{{et al.}}

\newcommand {\mm}[1] {\ifmmode{#1}\else{\mbox{\(#1\)}}\fi}
\newcommand{\GTO}        {\mm{\mathsf{GTOPO30}}}
\newcommand{\WarpX}        {\mm{\mathsf{WarpX}}}
\newcommand{\Nyx}    {\mm{\mathsf{Nyx}}}
\newcommand{\MICrONS}     {\mm{\mathsf{MICrONS}}}

\title{Distributed Augmentation, Hypersweeps, and Branch Decomposition\\ of Contour Trees for Scientific Exploration}

\author{\authororcid{Mingzhe Li}{0000-0003-0355-1919}, 
\authororcid{Hamish Carr}{0000-0001-6739-0283},
\authororcid{Oliver R{\"u}bel}{0000-0001-9902-1984},
\authororcid{Bei Wang}{0000-0002-9240-0700},
\authororcid{Gunther H. Weber}{0000-0002-1794-1398}}

\authorfooter{
\item Mingzhe Li is with the University of Utah.
  	E-mail: mingzhe.li@utah.edu
\item Hamish Carr is with the University of Leeds. E-mail: h.carr@leeds.ac.uk.
\item Oliver R{\"u}bel is with the Lawrence Berkeley National Laboratory. 
	E-mail: oruebel@lbl.gov.
\item Bei Wang is with the University of Utah. 
	E-mail: beiwang@sci.utah.edu. 
\item Gunther H. Weber is with the Lawrence Berkeley National Laboratory. 
	E-mail: ghweber@lbl.gov. 
}

\abstract{
Contour trees describe the topology of level sets in scalar fields and are widely used in topological data analysis and visualization.  
A main challenge of utilizing contour trees for large-scale scientific data is their computation at scale using high-performance computing. 
To address this challenge, recent work has introduced \emph{distributed hierarchical contour trees} for distributed computation and storage of contour trees. 
However, effective use of these distributed structures in analysis and visualization requires subsequent computation of geometric properties and branch decomposition to support contour extraction and exploration. 
In this work, we introduce \emph{distributed algorithms for augmentation, hypersweeps, and branch decomposition} that enable parallel computation of geometric properties, and support the use of distributed contour trees as query structures for scientific exploration. We evaluate the parallel performance of these algorithms and apply them to identify and extract important contours for scientific visualization.

}

\keywords{Contour trees, branch decomposition, parallel algorithms, computational topology, topological data analysis}

\teaser{
  \centering
  \includegraphics[width=0.95\linewidth]{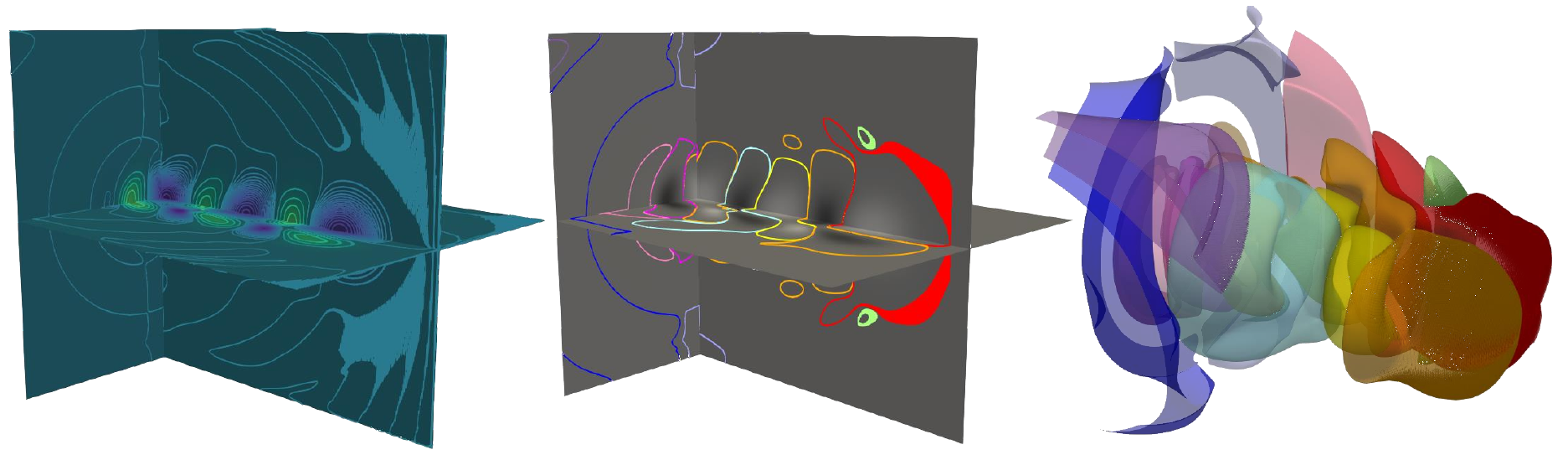}
  \caption{
    Our method applied to a 3D {\WarpX} laser-driven, plasma-based particle accelerator simulation dataset with a resolution of $6791\times371\times371$. We use the x-component of the electric field. Left: three 2D slices of the volume along different axes. Middle: 2D slices with the extracted contours on the slice. Right: Using parallel topological data analysis to extract and visualize 3D isosurfaces corresponding to the top-$11$ branches of the contour tree.
  }
  \label{fig:teaser-warpx}
}

\begin{document}
\maketitle

\section{Introduction}
\label{sec:introduction}

Topological data analysis abstracts features in scientific data mathematically with descriptors such as contour trees~\cite{BoyellRuston1963}, merge trees, Morse and Morse--Smale complexes~\cite{EdelsbrunnerHarerZomorodian2001}.
Although increasingly sophisticated tools have been developed to exploit these features, these tools often process topological descriptors using linear induction, giving rise to their serial computation (e.g.,~\cite{CarrSnoeyinkAxen2003,EdelsbrunnerHarerZomorodian2001,EdelsbrunnerHarerNatarajan2003}). 

Exascale computation requires scalable parallel analysis and visualization since it uses hybrid clusters with many machines, each with hundreds to thousands of cores, often on GPUs.
Since serial algorithms for topological descriptors are difficult to restate for this parallel computing environment, the advantages of topological data analysis have been hindered by the ability to compute the data structures and properties of topological descriptors at scale. 
To this end, there have been recent advances in the parallel and distributed computation of contour trees~\cite{MorozovWeber2014}, merge trees~\cite{MorozovWeber2013}, and Morse--Smale  complexes~\cite{ShivashankarNatarajan2012,ShivashankarMNatarajan2012,GyulassyBremerPascucci2019,CarrWeberSewell2021}.   

\update{In previous work, Carr~\etal~developed contour tree algorithms from serial~\cite{CarrWeberSewell2016} to shared-memory parallelism (SMP)~\cite{CarrWeberSewell2021}, and then added secondary computations such as acceleration structures~\cite{CarrRubelWeber2022a}, followed by geometric measures, branch decomposition, simplification and single-contour extraction~\cite{HristovWeberCarr2020}, all of which are available through the open-source VTK-m~\cite{MorelandSewellUsher2016} multicore toolkit.} 
For hybrid distributed-SMP, Carr~\etal~\cite{CarrRubelWeber2022b} reported a hybrid algorithm and data structure called the \emph{distributed hierarchical contour tree} (DHCT), implemented using VTK-m and the DIY toolkit~\cite{PeterkaRossGyulassy2011}.
However, effective use of these distributed structures in analysis and visualization requires substantial effort, in terms of geometric properties and branch decomposition; this is the focus of the current paper.  

We now report substantial extensions to the DHCT~\cite{CarrRubelWeber2022b}, in which we demonstrate how to augment a hierarchical contour tree with sufficient information to compute its branch decomposition and geometric properties over an entire cluster, and then demonstrate its use for analysis and visualization. 
Our contributions are as follows: 
\begin{itemize}[noitemsep]
\item	We introduce a distributed algorithm for augmenting a hierarchical contour trees. That is, we insert (regular) attachment points into the tree structure for later computations. 
\item We develop a distributed hypersweep algorithm for computing geometric properties of the contour trees. 
\item We describe a distributed algorithm for branch decomposition.
\item  We perform distributed extraction of individual contours from the branch decomposition for scientific visualization. 
\item   We include a systematic evaluation of the algorithm performance. 
\end{itemize}

\para{Outline.} 
Since much of this work depends on the prior state-of-the-art  for contour trees and their computation in serial and parallel, we review the technical background in \cref{sec:background}, in particular the distributed hierarchical contour tree~\cite{CarrRubelWeber2022b}.  
We introduce hybrid algorithms for augmentation, geometric measures, and branch decomposition in \cref{sec:augmentation} through \cref{sec:bd}, followed by distributed contour extraction in \cref{sec:isosurface}. 
We validate the correctness of the distributed contour tree computation in \cref{sec:validation} and discuss algorithm complexity in \cref{sec:complexity}. 
Finally, we demonstrate the full visualization pipeline in \cref{sec:results} and report performance results in \cref{sec:evaluation}, before stating conclusions and thoughts on future work in \cref{sec:conclusion}.

\begin{figure*}[!t]
\centering
\includegraphics[width=.99\textwidth]
{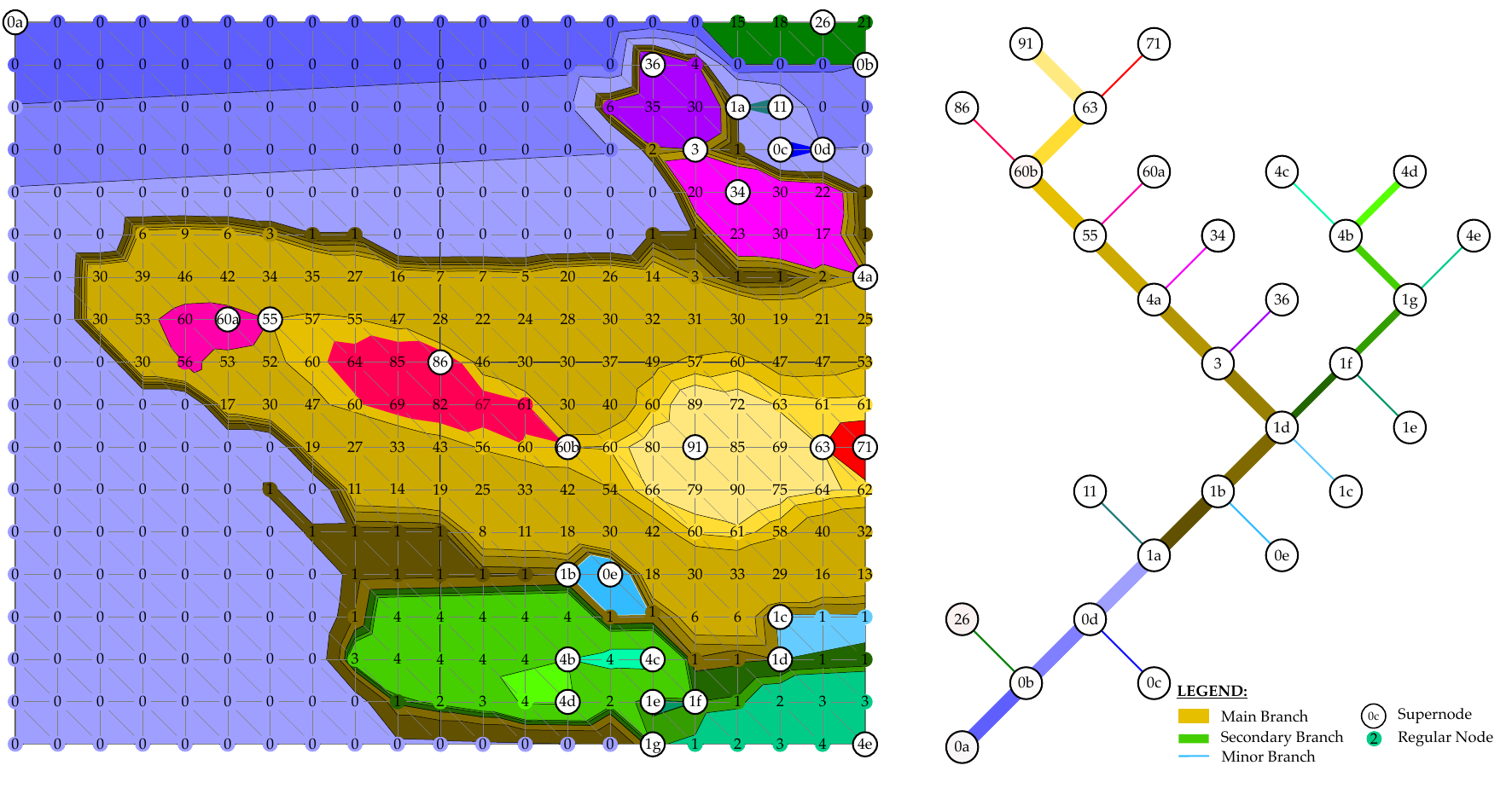}
\vspace{-6mm}
\caption{\update{A contour tree (right) and global topological zones (left) for Vancouver.  Data values on the underlying 2D  simplicial mesh are shown, with supernodes shown as hollow circles, and topological zones as colored bands matching the superarcs in the contour tree.  Letters indicate sort order due to the simulation of simplicity. The branch decomposition by approximated areas is shown on the right by line thickness, with $91-0a$ as the main branch, $4d-1d$ as the secondary branch, and all other superarcs forming individual branches (i.e,~minor branches).}}
\vspace{-4mm}
\label{fig:vancouver-unitary}
\end{figure*}

\section{Background}
\label{sec:background}

We first present contour trees mathematically (\cref{sec:contour-tree}), and then introduce serial algorithms for their construction, simplification,  and visualization (\cref{sec:ct-serial}).  
We then review the principal parallel algorithms (\cref{sec:ct-parallel}), and finally (\cref{sec:dhct}) the \emph{distributed hierarchical contour tree} (DHCT).

\subsection{Contour Trees}
\label{sec:contour-tree}

Given a scalar field $f: \Mspace \rightarrow \Rspace$ from a manifold $\Mspace$ to $\Rspace$, we define the \emph{level set} of an isovalue $h \in \Rspace$ to be $f^{-1}(h) = \{x \in \Mspace \mid f(x) = h\}$. We call connected components of a level set \emph{contours}~\cite{CarrSnoeyinkPanne2010}. 

Define an equivalence relation between $x, y \in \Mspace$, $x\,{\sim}\,y$ iff they belong to the same contour of $f$.
The quotient space of $\Mspace$ under this relation, $\Mspace/{\sim}$, is the \emph{Reeb graph} \cite{Reeb1946} of the function. 
\update{Informally, a Reeb graph is generated by contour contraction via the equivalence relation}.  If it is acyclic (i.e., it has no cycles), e.g.,~when the domain $\Mspace$ is simply connected, it is called a \emph{contour tree}~\cite{BoyellRuston1963}, denoted as $T_f(\Mspace)$. 

Contour tree vertices are called \emph{supernodes}, and are a subset of the critical points of $f$. The edges of a contour tree are \emph{superarcs}, which may be broken into \emph{arcs} by regular points called \emph{regular nodes}. A contour tree with added regular nodes and arcs is called \emph{augmented}.

We use the mapping $C: \Mspace \rightarrow T_f(\Mspace)$ from points in $\Mspace$ to the corresponding points in the contour tree and \emph{topological zones} for the inverse images under this mapping of superarcs, supernodes, or arbitrary subgraphs. 
\cref{fig:vancouver-unitary} shows a small example of a contour tree on a simplicial mesh, with colors indicating topological zones.  Where two supernodes have the same isovalue, letters are used to indicate the sort order induced by the simulation of simplicity~\cite{EdelsbrunnerMucke1990}.

\subsection{Serial Contour Tree Computations}
\label{sec:ct-serial}

The contour tree can be computed in serial by processing mesh vertices in a sorted order, tracking changes in contour connectivity \cite{KreveldOostrumBajaj1997}.  
A more efficient approach~\cite{CarrSnoeyinkAxen2003} (for all dimensions) sorts and sweeps through the mesh in each direction to compute \emph{merge trees}, capturing connectivity of super- and  sub-level sets of the form $\{x: f(x) \geq h\}$ and $\{x: f(x) \leq h\}$. 
The \emph{merge phase} of the algorithm then transfers leaf edges from the merge trees until the entire contour tree has been found.  
For non-simplicial meshes, the same algorithm requires a \emph{topology graph} that captures the essential topology of the mesh \cite{CarrSnoeyink2009}.

Geometric properties, such as \emph{contour length}, enclosed \emph{area} or \emph{volume} in 2D,  enclosed \emph{volume} or \emph{hypervolume} in 3D, as well as  mean, standard deviation, and root mean square (RMS), are functions of the isovalue \cite{BajajPascucciSchikore1997}, and can be computed by sweeping contours through the contour tree\cite{CarrSnoeyinkPanne2010} with three operations: \emph{update} at a regular vertex, \emph{inversion} as the sweep direction reverses, and \emph{combination} to reconcile functions for multiple contours at saddle points.

For regular grids, area (in 2D) or volume (in 3D) can be approximated by counting the vertices of the mesh in each zone \cite{CarrSnoeyinkPanne2010}, which is cheaper to compute than the exact value. Similarly, volume (in 2D) or hypervolume (in 3D) can be approximated by summing the data values.

In addition, the \emph{height} of a superarc or monotone path in the contour tree is the difference in isovalues between the endpoints of the superarc or path.  For merge trees, this height is the same as \emph{persistence}\cite{EdelsbrunnerLetscherZomorodian2002}, but for contour trees, this is not guaranteed~\cite{HristovCarr2021}.

The contour tree can be simplified by choosing a property as a measure of importance and removing the least important leaf iteratively until the tree is collapsed or it satisfies a desired condition such as size. As leaves are removed (or \emph{pruned}), critical points become regular, allowing superarcs to be coalesced into monotone paths called~\emph{branches}. 
Ultimately, when the entire tree has been pruned, a hierarchy of branches called a \emph{branch decomposition} is constructed \cite{PascucciColeMcLaughlin2003}.  
\update{\cref{fig:vancouver-unitary} shows the contour tree for a sample mesh, with the branch decomposition marked by varying the thickness of superarcs.}

\begin{figure*}[!t]
\centering
\includegraphics[width=.8\textwidth]{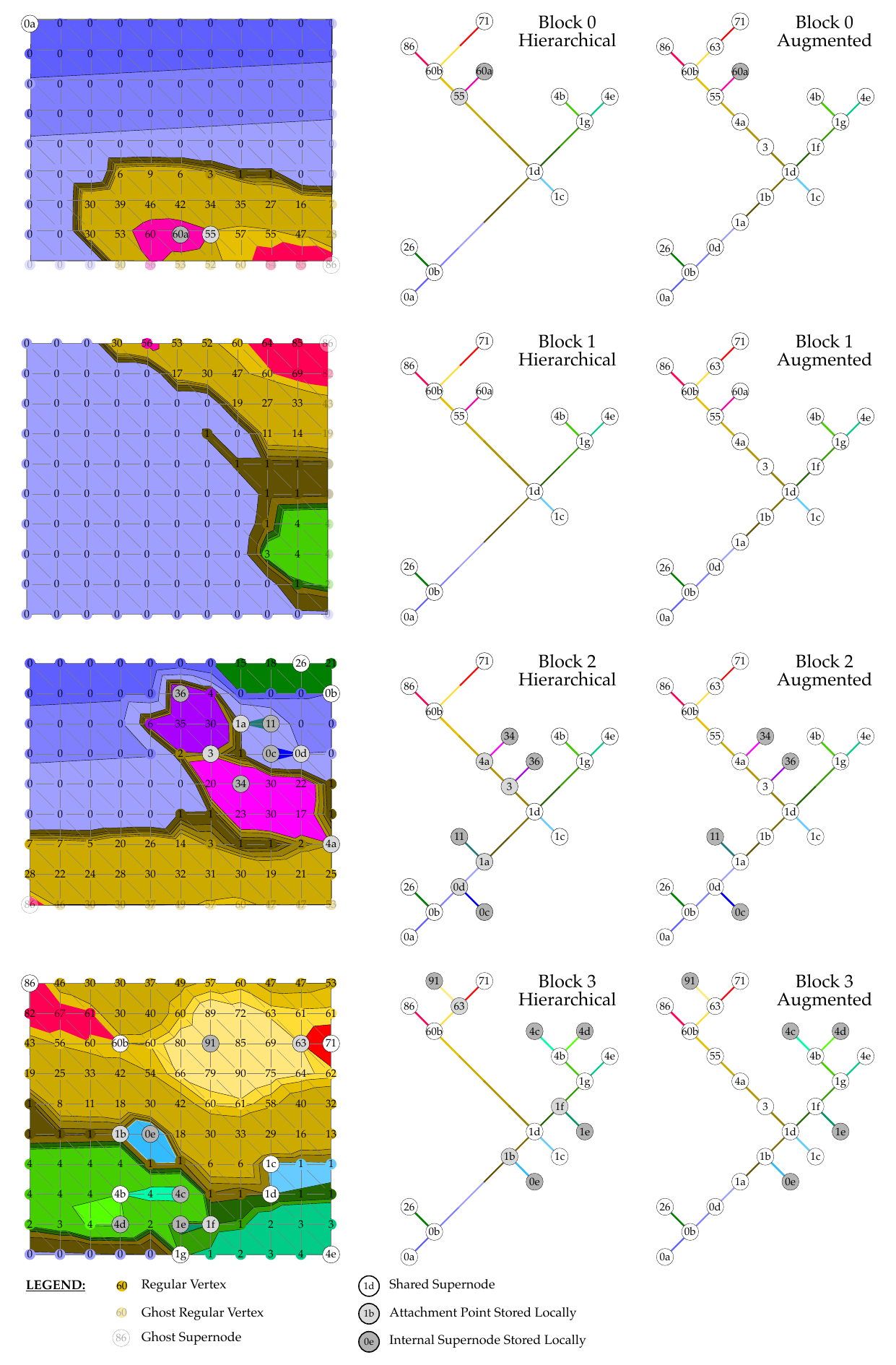}
\caption{\update{Distributed hierarchical contour trees for \cref{fig:vancouver-unitary}. Each block (rank) stores the subset of the global contour tree for all contours that pass through the block: the global contour tree is therefore the union of the individual trees. Prior to augmentation, subtrees restricted to a child block are not represented on other blocks. In the augmented version, attachment points for child subtrees are explicitly represented on all blocks sharing the superarc. The augmentation allows correct prefix scans but increases communication cost and memory footprints. Note that superarcs are assumed to be oriented towards the root $1d$, and are labelled by their source supernode.}}
\label{fig:vancouver-blockwise}
\end{figure*}

\subsection{Parallel Contour Tree Computation}
\label{sec:ct-parallel}

For parallel and distributed computation, we use the term \emph{block} as the unit to split the data, and the term \emph{rank} as the unit for the computational memory. In our experimental design, we always assign one data block to a rank as a convention to utilize the computational resources. These two terms are sometimes used interchangeably when there is no confusion from the context. We note that \emph{fan-out} refers to spreading a task to multiple destinations in parallel, and \emph{fan-in} does the opposite by sending multiple tasks to the same destination.

Pascucci and Cole-McLaughlin \cite{PascucciColeMcLaughlin2003} introduced a distributed computation of the contour tree by computing separate trees for individual blocks of a dataset.  
Gluing the trees from adjacent blocks together constructs a topology graph for the combined block, and the contour tree of the entire mesh can be computed with a recursive fan-in on a cluster \cite{PascucciColeMcLaughlin2003}. 
However, the computation stores the contour tree on the principal node, which is problematic for noisy data, where the contour trees can be linear in the input size.  
Moreover, this approach exploits only message-passing parallelism, rather than the shared-memory approaches typical of GPUs and individual nodes of a cluster.

Increasing core counts drives the adoption of shared-memory parallelism (SMP). 
Acharya and Natarajan introduced an approach that parallelizes the construction of topology graphs~\cite{AcharyaNatarajan2015} but uses serial computation of the merge and contour trees thereafter. 
The \emph{contour forest} introduced by Gueunet \etal~\cite{GueunetFortinJomier2016} breaks the mesh into segments by isovalue, computes separate contour trees, and glues them together at the boundaries.  
A subsequent  approach~\cite{GueunetFortinJomier2017} computes subtrees independently, using task-based parallelism to construct the tree a few edges at a time. 
Smirnov and Morozov~\cite{SmirnovMorozov2020} introduced a representation of merge trees called the \emph{triplet merge tree} that tracks the nesting of branches by propagating changes locally until the entire tree is computed,  supporting easy-to-parallelize algorithms in shared memory.  

Recently, Carr \etal~\cite{CarrWeberSewell2021} introduced the \emph{parallel peak pruning (PPP)} based on array-parallel operations.~PPP extends the approach by Acharya and Natarajan~\cite{AcharyaNatarajan2015} by constructing monotone paths from all vertices to extrema, and performing parallel pruning of all upper (or lower) leaves to construct a merge tree efficiently, and batching the final merge phase by alternating upper and lower leaf transfers.  

Subsequent work on PPP \cite{CarrRubelWeber2022a} described a data structure related to parallel rake-and-contract \cite{GibbonsCaiSkillicorn1994} tree acceleration called the \emph{hyperstructure} for efficient parallel operations on the entire contour tree.  In this work, monotone chains of superarcs called \emph{hyperarcs} are constructed pointing inwards a root node, sorting supernodes along each hyperarc to allow binary rather than linear search for locations in the tree. 
For any geometric property whose update rule can be converted to a prefix-sum operation along a hyperarc, this approach then enables parallel computation of geometric properties with \emph{hypersweeps}~\cite{HristovWeberCarr2020}.  
 
Hristov~\etal\cite{HristovWeberCarr2020} observed that SMP simplification or branch decomposition depends on the supernodes identifying which branch they belong to. In parallel, each supernode selects its \emph{best} ascending and descending superarc locally.  For height-based branch decomposition, parallel persistence is difficult to compute, and the min-max range in the subtree is therefore substituted. For region size (area/volume), however, the number of regular nodes in the subtree is easy to compute in parallel as an approximation of area (2D) / volume (3D). A modified pointer-doubling operation is then used to collect sets of superarcs into branches, using the maximal vertex in the branch as an identifier~\cite{HristovWeberCarr2020}.

To extract single contours for visualization, we note that all monotone paths (in a simplex) map to the corresponding monotone path between the minimum and maximum of the simplex in the contour tree \cite{WeberDillardCarr2007}. 
All simplices are tested in parallel to see whether their path in the tree passes through the point in the tree representing a desired contour.
The corresponding cells's contour is then extracted if needed, using the hyperstructure for acceleration \cite{HristovWeberCarr2020}. 
Last, but not least, these operations allow users to select the $k$ most interesting contours in the dataset, extract and render them \textsl{in situ} \cite{HristovWeberCarr2020}, and store them using the CinemaDB system \cite{AhrensJourdainOLeary2014} to enable limited \textsl{post hoc} interaction.

\subsection{Distributed Hierarchical Contour Trees}
\label{sec:dhct}

\update{With efficient SMP algorithms in hand, we turn to the problem of hybrid distributed computation.} 
While a distributed algorithm was  proposed~\cite{PascucciColeMcLaughlin2003}, it stored the entire contour tree on a single rank (i.e.,~block). 
For noisy data, this approach not only drove up the communication cost, but also had a large memory footprint on that rank.  Carr~\etal\cite{CarrRubelWeber2022a} therefore proposed a method for distributed contour tree computation that reduces distributed memory footprint and communication cost, as well as exploiting local SMP parallelism using the PPP algorithm.

The approach by Carr~\etal\cite{CarrRubelWeber2022a} is based on the observation that some topological zones are in the interior of a single block, such as the light yellow zone that maps to the superarc $91 - 63$ in Block 3 of \cref{fig:vancouver-blockwise}. This zone (and superarc) is only stored on the rank where the block is held.  These zones and superarcs are removed from the tree, and the existing distributed algorithm~\cite{PascucciColeMcLaughlin2003} applies recursively: at each stage of the fan-in, the blocks for two or more children are combined, the shared contour tree is computed, and zones interior to the combined block are suppressed. 

At the end of the fan-in, this method has computed the shared contour tree for all blocks: for convenience, we  compute the contour tree on all ranks rather than have a separate fan-out stage to communicate it. This tree, however, does not hold the superarcs that are suppressed during the fan-in, so each rank now re-inserts them one level at a time in a layered version of the hyperstructure called a \emph{hierarchical contour tree}.

At the end, all superarcs are correct in at least one block, and the union of all hierarchical contour trees is the correct \emph{distributed hierarchical contour tree} (DHCT), \update{shown in \cref{fig:vancouver-blockwise} for a four-block subdivision of a dataset.} 

Both the DHCT and the hyperstructure are built iteratively, with early iterations building the center of the tree and being treated as more \emph{senior} than later iterations.

The aim is to minimize the communication cost and memory footprint. Since the removal of a superarc such as $91-63$ in \cref{fig:vancouver-unitary} leaves supernode $63$ structurally redundant, we suppress not only superarc $91-63$ but also supernode $63$.  This approach is more efficient, but it comes at a cost for the secondary computations, where the hypersweep computation~\cite{HristovWeberCarr2020} depends on the correct order of supernodes along each superarc.  As an example, if we look at \cref{fig:vancouver-blockwise}, we see that $1a,0d$ are represented only in the tree for Block $2$ and $1b$ in the tree for Block $3$, but any computation that depends on knowing the correct order of $1a,0d$, and $1b$ will fail unless all ranks are aware of these supernodes.

Since the branch decomposition in particular depends on the correct ordering of superarcs along a hyperarc, we must augment the DHCT with these additional nodes before proceeding. This leads to our first contribution, detailed in \cref{sec:augmentation}.
\section{Hierarchical Augmentation}
\label{sec:augmentation}

\begin{figure*}[!t]
\centering
\includegraphics[width=\textwidth]{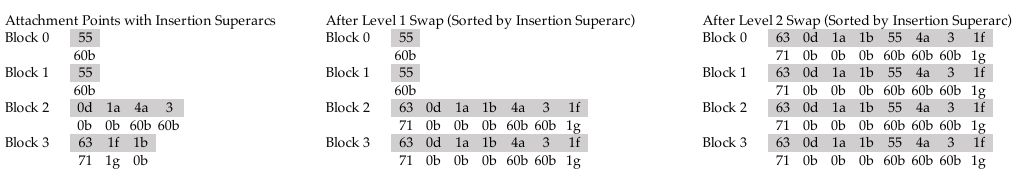}
\vspace{-8mm}
\caption{\update{Swap sequence for augmentation.  Ranks swap their list of attachments with their partners until all ranks have the full list of attachment points to insert along each shared superarc on the rank.}}
\label{fig:superarcSwap}
\vspace{-4mm}
\end{figure*}

A DHCT keeps features locally to minimize communication, recording the saddle or \emph{attachment point} at which each local branch attaches to a superarc stored at a higher level of the hierarchy. 
\update{Since the zone is internal to the block, other blocks do not need the attachment point even if it is on the boundary between blocks.  
The DHCT therefore adopts a \emph{lazy} insertion policy by storing the attachment points locally and inserting them only when needed for local computations.}

Hypersweeps, however, rely on prefix operations for efficient parallel computation and need the exact order and values of all supernodes on a hyperarc to propagate subtree computations correctly.  Worse, when computing a branch decomposition, each rank needs to identify the best upwards and downwards branches for each supernode, and the best upwards or downwards branch may be located on another rank.

In order to perform distributed hypersweeps and branch decomposition, we need to insert attachment points from the hierarchical contour trees so that all ranks share the information necessary. We note that this process involves augmenting the contour tree with a set of points, and that the same process can be used for any arbitrary set of insertions.

\update{Attachment points are always existing vertices in the topology graph and supernodes in the block's hierarchical contour tree, but may not be supernodes in another block's tree.} 
Thus, in order to perform hypersweeps, we must collect the set of points to be inserted on each shared hyperarc, and then construct the sequence of supernodes and superarcs in the hyperstructure following the process described in~\cite{CarrRubelWeber2022a}.

Since the DHCT uses a specific fan-in sequence of block combination, augmentation follows the same sequence. For each level of the hierarchy, each rank exchanges with its partner all attachment points along every superarc at that level or higher. As we fan in, these exchange sets grow until each node has the full set of attachment points for a given superarc at the fan-in level to which the superarc belongs. 

\update{After the fan-in is complete, a fan-out starts, breaking old superarcs into sequences of new superarcs at each level.} This approach preserves the properties of the hyperstructure that superarcs (along a given hyperarc) are stored (in sequence) in the arrays, and that superarcs shared between blocks always share ID numbers. We end up with a new  valid DHCT that explicitly represents all attachment points rather than storing them locally for lazy insertion. 

We note that, except for a change in vertex numbering, the hyperstructure is unaffected by the augmentation, except that the attachment points no longer have hyperarcs representing their connection to their parent superarc.  As a result, the first stage of the augmentation is to copy the hyperstructure, leaving the vertex renumbering until the end. 

The above algorithm can be expressed as follows:
\begin{itemize}[noitemsep,leftmargin=*]
	\item	For each rank
    \begin{enumerate}[noitemsep,leftmargin=*]
    	\item	Initialize working arrays for swaps
        \begin{enumerate}[noitemsep]
            \item	Compress the list of supernodes to attachment points only
            \item	Copy supernodes into a swap array segmented by level
        \end{enumerate}
        \item	For each round of fan-in
        \begin{enumerate}[noitemsep]
	   	   \item	Swap the array prefix with the partner for shared superarcs 
    	   	\item	Update the local list of attachment points per superarc
	   \end{enumerate}
	\item	Suppress duplicate attachment points
	\item	For each round of fan-out
        \begin{enumerate}[noitemsep]
            \item	Copy supernodes from DHCT to the end of new DHCT
            \item	Add attachment points to the end of new DHCT
            \item	Sort supernodes along hyperarcs for the round
            \item	Set superarc targets for the round	
        \end{enumerate}
    \item	Update the hyperstructure with new supernode IDs
    \item	Copy all remaining regular nodes to the new DHCT	
    \end{enumerate}
\end{itemize}

\cref{fig:vancouver-blockwise} shows the result of this computation for the dataset from \cref{fig:vancouver-unitary}. The first column shows the data blocks for reference, and the second shows the DHCT before augmentation, with attachment points in grey, and the third shows the DHCT after augmentation, with all attachment points now explicit in white, as they have been shared with all other blocks storing the parent superarcs.

\cref{fig:superarcSwap} shows the swap sequence of attachment points (gray) and their insertion superarcs (white). 
We do not require any level $0$ swaps, as insertions happen only at higher levels. 
\update{The level $1$ swap exchanges between Blocks $0$ \& $1$ and between Blocks $2$ \& $3$, building up lists of attachment points to be inserted for shared superarcs $71-60b, 0b-1d, 60b-1d$, and $1g-1d$; note that we have sorted them for clarity, but the code defers the sort until the full set is assembled.}
At level $2$, Blocks $0$ \& $2$ swap, as do Blocks $1$ \& $3$, until all ranks have the full set of attachment points to be inserted along the top-level superarcs.

\update{During the fan-out, we reconstruct the DHCT to produce the new versions shown in \cref{fig:vancouver-blockwise}. We can see that nodes $0d, 1a$, and $1b$ will all be inserted along $0b-1d$ to produce new superarcs.}



\section{Distributed Hypersweeps}
\label{sec:hypersweeps}

\begin{figure*}[!t]
\centering
\includegraphics[width=\textwidth]{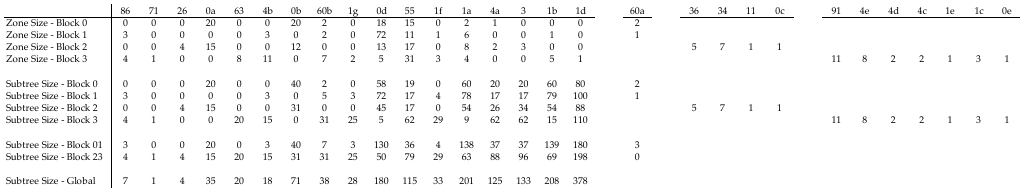}
\caption{\update{Distributed hypersweep for subtree node count.  Each block uses prefix sums to count regular nodes in each zone, omitting nodes shared with blocks with higher block ID (white lettering in \cref{fig:vancouver-blockwise}).  Each block then performs a hypersweep to compute the contribution to the subtree size. Blocks then sum region sizes in the same fan-in used for the trees.}}
\vspace{-4mm}
\label{fig:region-sums}
\end{figure*}

As described in \cref{sec:augmentation}, we augment the DHCT to ensure that all supernodes necessary for prefix operations are correctly represented on all relevant ranks. We can now describe how to distribute the hypersweep computation, using volume (subtree size) as the running example.

\update{Hypersweeps work for any associative property~\cite{HristovWeberCarr2020}, and we add a distributed stage. For the hypersweep, we compute the property separately for each block, and then use a fan-in computation to combine them.}  \cref{fig:region-sums} illustrates this approach for the example shown in \cref{fig:vancouver-blockwise}, using the regular node count as an approximation of area.

We compute the number of regular nodes in each zone for each block, noting that boundary nodes are represented on multiple blocks. We adopt a simple rule: each such node belongs to the block with the higher ID, so each node is counted only once. 
\update{In \cref{fig:vancouver-blockwise}, boundary nodes (in the domain) are shown as ghost nodes (faded) in all blocks except the one to which they belong.}

Counting is then performed locally as before\cite{HristovWeberCarr2020}, and is shown in the first four rows of \cref{fig:region-sums}. We separate the zones by level of the hierarchical contour tree, so that all shared zones are listed first.

Once we have computed the local zone size for each block, we perform a local hypersweep\cite{HristovWeberCarr2020} to get the local subtree size for each block, shown in the next four rows.  

This approach can therefore be summarized as:

\begin{enumerate}[noitemsep]
	\item	Assign each vertex to belong to one rank only
	\item	Compute superarc measures locally
	\item	Use the local hypersweep to compute the subtree measure
	\item	For each round of fan-in
	\begin{enumerate} [noitemsep]
		\item	Swap this round's measure with the partner
		\item	Combine the partner's measure with its own measure
	\end{enumerate}
\end{enumerate}

We note that this exchange involves the same amount of data as the initial construction of the DHCT in the first place: we will see in \cref{sec:evaluation} that the practical cost of data exchange is dominated by the cost of computing the augmented tree.

\section{Distributed Branch Decomposition}
\label{sec:bd}

Once the desired geometric measure has been computed, the next step is to compute a distributed branch decomposition.  We recall from previous work \cite{HristovWeberCarr2020} that each supernode chooses the best ascending and descending branch locally.  Once the best up/down branch is known, a modified pointer-doubling stage collects all branches in parallel, using the highest vertex on the branch as the representative of the set.

Again, the distributed version needs some modification, as it is possible for the best ascending branch to belong to another block.  We resolve this by having an initial pass in which all blocks separately compute the best ascending branch and exchange with their partners in the hierarchy, updating the best branch after each swap until the top of the hierarchy is reached.  At this point, all blocks agree on which branch is preferred, and can perform the pointer-doubling to determine the set of supernodes known to that block that belong to the branch.  

However, in doing so, we observe that if the highest vertex on the branch is not shared, we will not agree on the correct representative for the branch.  We therefore choose the most senior supernode on the branch (i.e., the one closest to the root of the hyperstructure), which is guaranteed to be shared between all blocks that include the branch even partially, always using the global vertex ID as an identifier rather than a local supernode or regular node ID.  We note that this sense of seniority is \emph{specific} to the hyperstructure and DHCT, and is \emph{not} the same as whether a branch is main or secondary.

We also observe that since the branch decomposition does not necessarily follow the hyperstructure, a branch may not have all of its superarcs oriented in the same direction. We resolve this by orienting all superarcs toward the \update{seniormost} end (i.e., the one closest to the root of the hyperstructure).  In other words, instead of the pointer-doubling collapsing the branch toward the topmost supernode, it collapses the branch toward the root. Once we realize this, however, we see that the superarcs must be oriented toward the end closest to the root.  This is easy to test, and results in the pointer-doubling operating over a forest of directed edges, where each component has a well-defined root.

Since we now know the branch of all superarcs in each block, we sort all superarcs by their branch, their scalar value, and their global vertex ID (for the simulation of simplicity) to get the local upper- and lower-end supernodes of the branch. 
Following the sort, we need to exchange the branch ends across blocks to agree on the global upper- and lower-end supernodes of each branch. With the branch end information, we can compute the volume of branches based on the subtree volume of the superarc at the saddle end of the branch. 

The steps are summarized as follows:

\begin{enumerate}[noitemsep]
	\item	Choose the local best up/down for each supernode 
	\item	For each round of fan-in
	\begin{enumerate}[noitemsep]
		\item	Swap the best up/down and volumes with the partner	
		\item	Choose the best of own/partner's up/down
	\end{enumerate}
	\item	Build branches in the local version of tree
    \item	Sort superarcs to get local branch end supernodes
    \item	Exchange branch end information to agree on the global upper/lower end of branches
    \item	Compute the branch volume using the ending superarc volume
\end{enumerate}

%
%
%

\section{Distributed Contour Extraction}
\label{sec:isosurface}

Now that we have the volume of some branches in each block, we select the top $k$ branches by volume for each block, excluding the main branch, of which both ends are leaf nodes (which will be included at a later step). Then, we exchange this information to agree on the global top $k$ branches. We extract contours based on these branches.

\begin{figure}[!ht]
\vspace{-4mm}
\centering
\includegraphics[width=0.98\columnwidth]{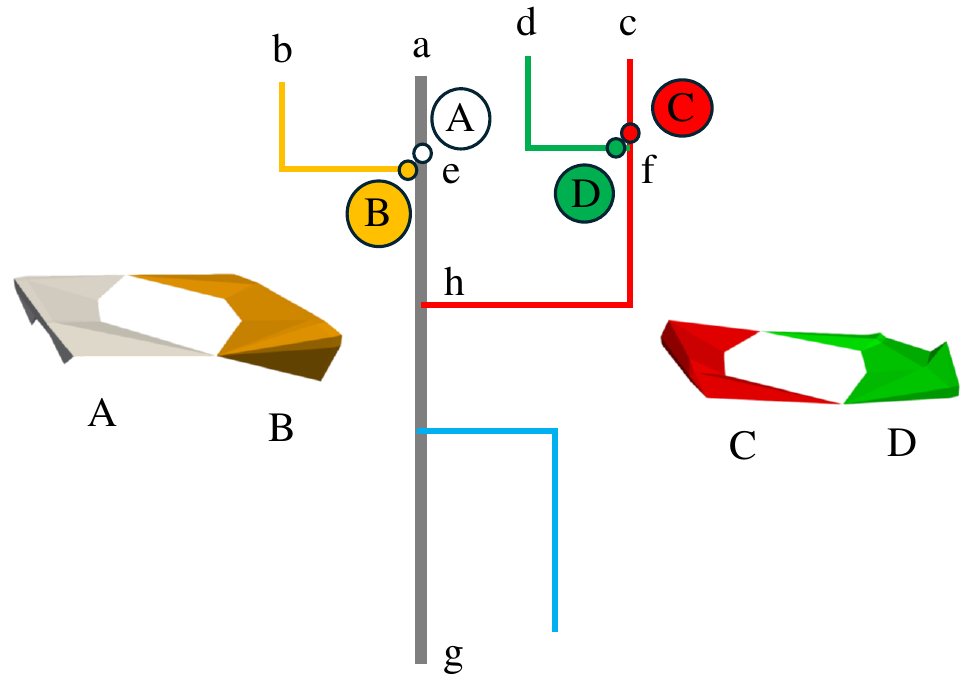}
\vspace{-4mm}
\caption{An example for contour extraction. Critical points are labeled with lower case letters, and contours with upper case letters.}
\label{fig:isosurface-example}
\vspace{-2mm}
\end{figure}

\update{Given a branch decomposition, we extract a representative set of contours.  Since 3D contours along the same branch normally occlude each other, we choose  \emph{local contours}~\cite{CarrSnoeyinkPanne2010}: a set of $k$ superarcs (or branches) with one contour on each.  Although any $k$ branches may be chosen, the normal choice is the $k$ most important branches in the tree.} 

\update{Once we have selected which branches to place contours on, we must choose an isovalue.  Placing a contour at the leaf end of the branch is unlikely to be useful, as it will often be so small as to be invisible, so the obvious choice is to take an isovalue at the saddle end of the branch, where the feature has its largest extent.  For example, in \cref{fig:isosurface-example}, we set $k = 4$. Then, placing a contour at the leaf end of branch $d-f$ is not a good choice, so we place a contour at $D$, just above the saddle.  On the main branch, which has two leaf ends, we arbitrarily select one as the "leaf" for this purpose, often the maximum.}

\update{If one of the branches chosen is the child of another, the parent's contour is likely to occlude the child's, and we instead choose the value just before the highest (or lowest) saddle at which any child joins the branch.  An example of the contour choice can be seen at $C$ in \cref{fig:isosurface-example}, where the child $d-f$ connecting to branch $c-h$ at $f$ means that we need to place the contour at $C$, just above $f$, rather than just above $h$.} 

\update{
To find the desired isovalues for each contour, we construct a \emph{branch decomposition tree} to capture the parent-child relationships, which are represented only implicitly in the DHCT.  For this step, we first identify the $k$ most important branches, and then their parents.  We then take the \emph{branch saddle}: the saddle end of a given branch, and retrieve the corresponding branch ID to find the parent: a simple lookup.  Once this is done, local analysis identifies the isovalue chosen for each branch.}

\update{
In distributed computation, the process is less easy.  Each block selects the $k$ most important branches that it stores and the importance measures for them, and then swaps with a fan-in partner to get at most $2k$ branches.  The block then reduces this to $k$ again, and swaps with the next partner, and so on, until all blocks determine which $k$ branches are most important, and the balance of the computation is then the same.}

\update{Finally, we extract the actual contours in each block, in a distributed version of the method of Hristov~et~al~\cite{HristovWeberCarr2020}. 
For each tetrahedron (in 2D, triangle) in the input mesh, we use Marching Cells as usual to extract a local contour fragment~\cite{BL03_substitopes, BWC04_anyd_isosurface}, and then consult the hyperstructure to find the superarc and branch to which the fragment belongs: if the branch of the superarc is not 
 as desired, the fragment is discarded. }

Overall, the implementation of this phase is then:

\begin{enumerate}[noitemsep]
    \item	Locally select branches to extract contours based on measures
    \item   Exchange the selected branches for the globally selected branches
    \item   Compute the relation between top-volume branches in local blocks, including the main branch
    \item   Merge the local branch relation for the global branch relation
    \item   Determine the (branch, isovalue) pair for each contour
	\item   For each branch and an isovalue
    \begin{enumerate} [noitemsep]
        \item Compute the active cells and their superarc
        \item Add the cell to the result if its superarc is on the branch
    \end{enumerate}
\end{enumerate}
\section{Correctness Validation}
\label{sec:validation}
 
For the DHCT, validation of correctness becomes increasingly difficult in the development process. Throughout the development of PPP, Carr \etal~validated results against a modified version of the serial contour tree algorithm~\cite{CarrSnoeyinkAxen2003}, using text files to ensure that the same tree was computed in parallel. As an additional check, the PPP code was implemented outside VTK-m as well as inside, and the results were compared against each other.

\update{For a distributed contour tree, three versions were used:} a monolithic version in which the data for all ranks are stored in shared memory simulating MPI-style communication, the second a full distributed MPI implementation, \update{and the third the VTK-m code, which was based on the standalone MPI implementation instead of the full MPI version.}

Our overall goal was to confirm that the distributed representation produces the same result as the serial implementation, despite the full contour tree being distributed across many compute nodes.  Moreover, where even the largest single compute node has insufficient memory for the entire tree, it should still be possible to test whether runs on different numbers of compute nodes at least achieve the same result.  Therefore, while we implemented the ability to combine the trees onto a single compute node in VTK-m, we do not rely on this  implementation for validation.  Instead, each rank saves its hierarchical contour tree to local disk, representing each supernode as a tuple (top, bottom, value, global) on each superarc to which it belongs, using global IDs for all vertices.  These files are concatenated and sorted externally, and duplicates are suppressed to produce a canonical list of all supernodes on each superarc for checking against either the serial code or runs with different numbers of ranks.

While this approach was successful early on, later stages required specialized external code that assembled the correct solution from the partial files, in particular when validating the branch decomposition.  Eventually, we also implemented a routine in VTK-m for testing purposes that assembles the entire tree on a single node, but observe that for the biggest files, the external approach will still be needed.

\section{Complexity}
\label{sec:complexity}

In this section, we discuss the asymptotic complexity before the runtime performance in~\cref{sec:results}.
\update{Let $N$ be the number of mesh (regular) vertices and $t$ the size of the contour tree. 
Let $B$ be the number of branches in the branch decomposition, and $k$ the number of branches from which we extract contours. 
$N_b$ denotes the number of blocks in the computation, and $r = \lg{N_b}$ denotes the number of rounds of distributed communication. 
These parameters reflect the global sizes for the full mesh, rather than per-block sizes.  
However, since the later stages of the fan-in could  potentially be global, the provable bounds are  based on the global sizes, and as a result, the bounds are fairly loose in practice.  
To these bounds, we observe that $k \leq B < t$.}

\para{Cost of augmentation.}~Since augmentation follows the same sequence of swaps as the computation of the DHCT, we need $r$ swaps to exchange attachment points.  Each swap has a total communication cost of $O(t)$, a time cost of $O(\lg{t})$, and a total work cost of $O(t \lg{t})$ to combine lists of supernodes, sort them, and reinsert them.  However, the practical bound on the number of supernodes exchanged is roughly linear in the size of the boundary, as only attachment points on the boundary were needed in the fan-in~\cite{CarrRubelWeber2022b}.  Barring W-structures~\cite{HristovCarr2021}, this practical bound can be proven, and even in the presence of the W-structures, it is tight in practice: for 3D data, the communication cost is thus bounded by $O(N^{2/3})$ in each round. Since this cost is similar to that needed for compute codes that exchange ghost zone information in each iteration, we feel that further optimization is  unnecessary.

Since we augment with non-boundary supernodes, this bound relaxes, and the worst-case scenario occurs when the contour tree consists of a single major branch, with $O(N)$ short side branches, evenly distributed between all ranks.  For this, there is a trivial tight bound of $\Omega(r t)$ for augmentation, geometric measure computation, and branch decomposition.  However, this behavior is not typical, and we focus instead on the empirical efficiency. 

\para{Cost of branch decomposition.} The cost of branch decomposition is similar to the augmentation, since further swaps are required of data relating to the augmented shared structures in the DHCT.  The overall cost is of at most $O(t \lg{t} r)$ work, $O(\lg{t} r)$ time, and $O(r t)$ communication, but the empirical bound is much lower.

\para{Cost of contour extraction.}~Choosing the $k$ best branches among $B$ branches requires a sort by importance, taking $O(B\lg B)$ work in $O(\lg B)$ time.  Since each rank operates independently, the communication necessary to agree on the $k$ most important edges requires $O(rk)$ communication. In practice, $k$ is usually small (e.g., $\leq 100$) due to the limitation of visual complexity and data memory.

\para{Summary.}  Since the principal limitation in distributed computation is communication cost, we expect that for the algorithms that depend on augmenting the DHCT, the tree size $t$ will be between the $O(N^{2/3})$ possible for the DHCT computation and the $O(N)$ worst-case augmented DHCT.  Thereafter, volume computation and branch decomposition will scale with the augmented tree size, while contour extraction will scale with the data size and the number $k$ of contours to be extracted.

\section{Results}
\label{sec:results}

We apply our algorithms in practice for scientific visualization, in particular, contour (isosurface) extraction (\cref{sec:result-isosurface-extraction}),  and evaluate their parallel performance (\cref{sec:evaluation}). 
We begin by providing an overview of our experimental design, including the implementation, datasets, compute resources, and parameter configurations.

\para{Implementation.}~Our implementation is based on the DHCT implementation in VTK-m~\cite{MorelandSewellUsher2016}. 
We extended the \texttt{ContourTreeUniformDistributed} filter to compute the augmentation and compute the branch volumes with hypersweeps in the DHCT. We also added \texttt{DistributedBranchDecomposition} and \texttt{SelectTopVolumeContours} filters to compute the branch decomposition and to select the top $k$ contours by volume, respectively.

VTK-m performs on-node SMP parallel computation exploiting OpenMP or CUDA or other mechanisms as appropriate, but does not use distributed parallelism.  We therefore added the DIY~\cite{morozov2016block} block-parallel library for distributed computation. 
Our implementation will be made available in the \href{https://gitlab.kitware.com/vtk/vtk-m}{VTK-m} repository in the near future.

\para{Datasets.}~We experiment on four large-scale datasets: {\GTO}, {\WarpX}, {\Nyx}, and {\MICrONS}. 

{\GTO}~\cite{cisl_rda_ds758.0} is an open-source 2D global terrain dataset, with 30 arc-second grid spacing (c. 1km) and a resolution of $21601\times43201$. 

{\WarpX}~\cite{FedeliHuebl2022WarpX} is the x-component of the electric field in a plasma laser-driven acceleration simulation, with asymmetric resolution, and an overall size $6791\times371\times371$.

{\Nyx} is a cosmological simulation of particle mass density\cite{Biwer2019}, based on LBNL compressible cosmological hydrodynamics simulation code ~\cite{AlmgrenBellLijewski2013} that solves flow equations in an expanding universe. We use dark matter density as the scalar field at a resolution of $1024 \times 1024 \times 1024$.

{\MICrONS}~\cite{ConsortiumBaeBaptiste2021} is a subvolume of $1024^3$ from Electron Microscopy (EM) image data of a P60 mouse cortex with a volume of $1.4$mm$\times 0.87$mm$\times0.84$mm from BossDB~\cite{bossdb}.

Both {\Nyx} and {\MICrONS} datasets have complex topology, whereas the {\WarpX} dataset has a relatively clean topology.

\para{Hardware configurations.} 
All experiments are performed on Perlmutter, a Hewlett Packard Enterprise Cray EX supercomputer with 3,072 CPU-only and 1,792 GPU-accelerated nodes at the National Energy Research Scientific Computing Center (NERSC). Each CPU-only node has two 2.45 GHz (up to 3.5 GHz) AMD EPYC 7763 (Milan) CPUs with $64$ cores per CPU and $512$ GB of DDR4 memory in total. Each physical core is equipped with two hardware threads (logical CPUs). 

\para{Parameter configurations.} We chose $k=10$ when selecting the top $k$ branches by volume for contour extraction. That is, we visualize contours  from at most $11$ branches, according to the contour extraction strategy (\cref{sec:isosurface}).
\update{An extensive parameter sensitivity analysis for the parameter $k$ is provided in the supplement.}
We use OpenMP~\cite{dagum1998openmp} for thread parallelization.

\para{Visualization generation.}~Visualizations are generated using ParaView 5.11.1~\cite{AhrensGeveciLaw2005} with VTK~\cite{VTK06}, with contours colored by the branch they belong to, and at most $11$ contour colors. We note that a high-volume branch may correspond to the background, with the isovalue choice rule resulting in a contour that is not visible in the visualization. For example, in the {\Nyx} dataset, the background contains vertices at zero density in the density field; in the {\GTO} dataset, the background contains vertices valued at $-9999$ that represent the ocean. These background volumes are omitted in~\cref{sec:result-isosurface-extraction}.

\begin{figure}[!ht]
\centering
\includegraphics[width=\columnwidth]{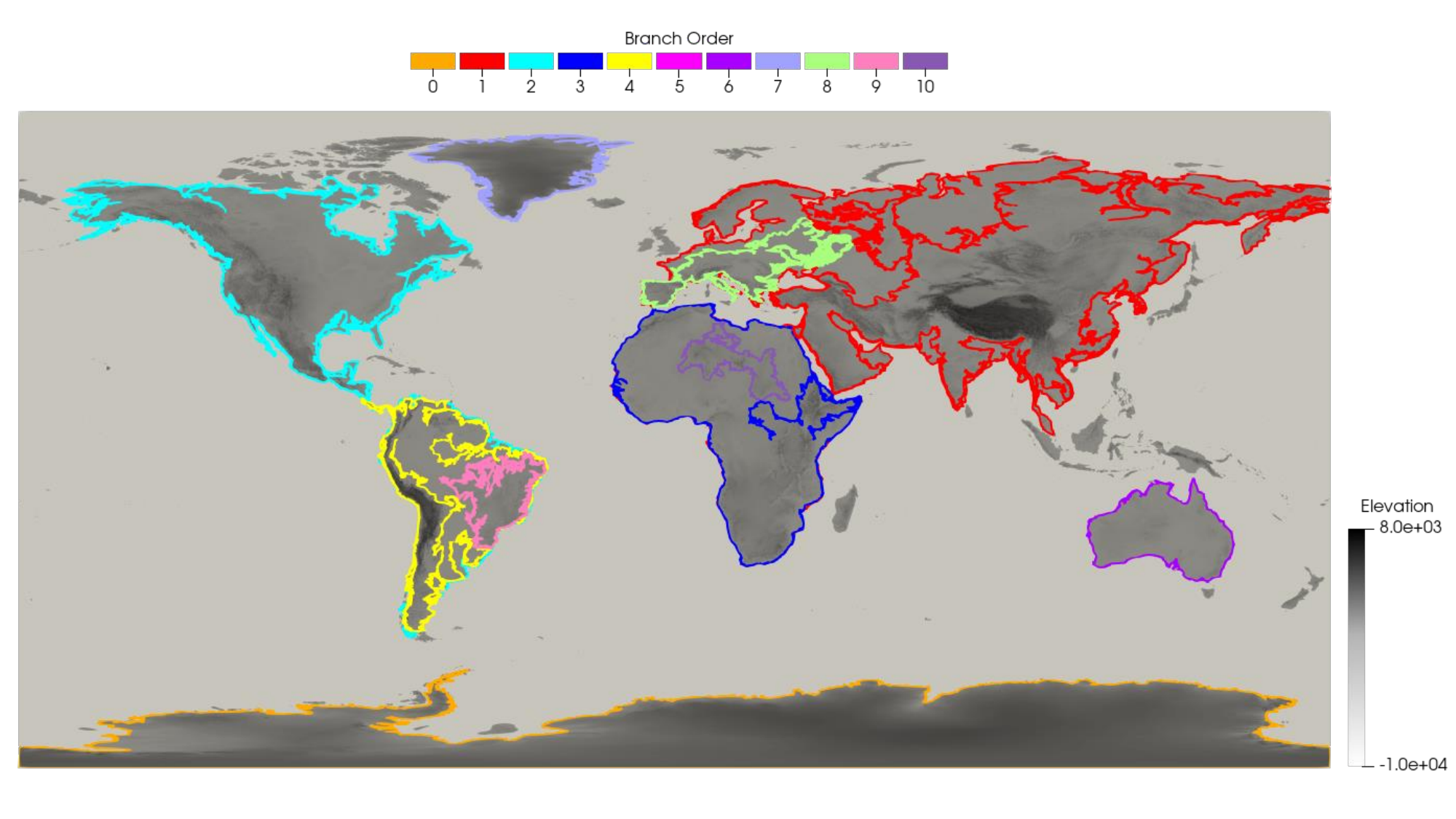}
\vspace{-8mm}
\caption{{\GTO} dataset. The contours shown enclose major regions and Greenland, as well as large highlands such as in Brazil.}
\vspace{-4mm}
\label{fig:gtopo-isosurface}
\end{figure}

\begin{figure}[!ht]
\centering
\includegraphics[width=0.8\columnwidth]{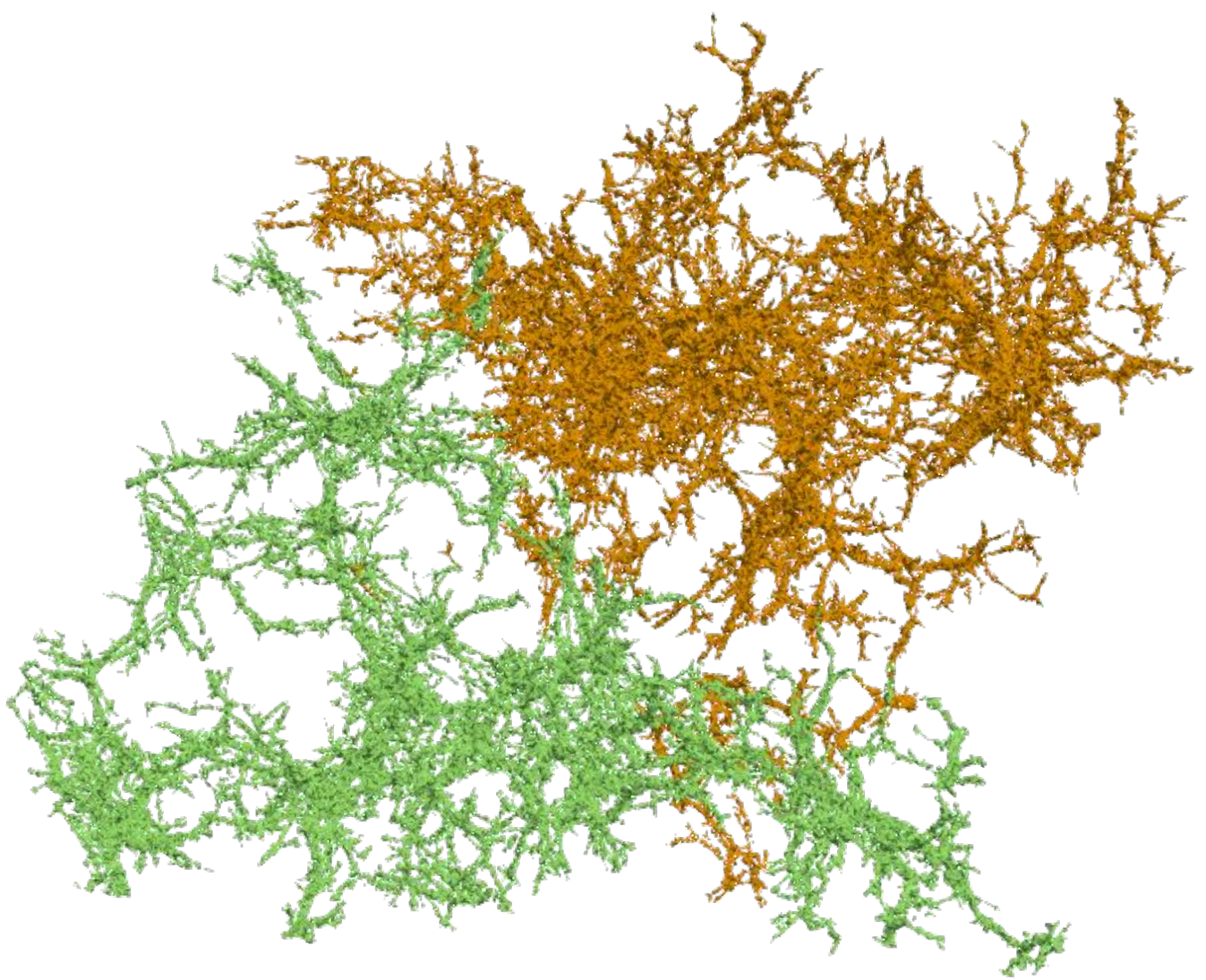}
\vspace{-2mm}
\caption{{\Nyx} dataset. The contours highlight the filaments in the Universe.}
\vspace{-4mm}
\label{fig:nyx-isosurface}
\end{figure}

\subsection{Contour Extraction}
\label{sec:result-isosurface-extraction}

We apply our distributed algorithms to identify and visualize relevant contours with large-scale volumes for the {\GTO}, {\WarpX}, and {\Nyx} datasets. 

\cref{fig:gtopo-isosurface} shows the elevation field of {\GTO} highlighting the top-ranked contours, which broadly depict the topologically prominent regions that map to the continents and Greenland, except South America, where the Brazilian Highland appears as a separate contour. Since {\GTO} is a 2D projection of the globe, the areas are distorted, so Antarctica is reported as the largest area (most important branch) in orange, followed by Eurasia in red, and America in cyan.  

\cref{fig:teaser-warpx} shows the top contours by volume for the {\WarpX} dataset. The 2D slices on the left demonstrate the characteristics of the scalar field, showing the positive and negative charge regions of the induced plasma wave shifting along the center x-axis. The selected contours  on the middle and right show the smooth boundaries separating the positive and negative volumes.
Defining these volumes allows for subsequent calculation of acceleration gradients (i.e., the sum of all values enclosed by the contour) and provides insight into the overall structure of the field. 
We can also observe important details from the contour: the high density of the red contour indicates the increasing frequency of the wave shift in the front of the acceleration direction.

In the visualization of {\Nyx} (\cref{fig:nyx-isosurface}), we see contours on only two branches due to the large background volumes (at zero density). The extracted contours, however, highlight the high-density filamentary structures in the Universe.

\subsection{Runtime Performance and Strong Scaling}
\label{sec:evaluation}

Our main evaluation is performance, focusing on strong scaling on CPU-only nodes, by considering scalability at a fixed problem size with different numbers of nodes for all four datasets. We also vary the number of ranks per node to evaluate the best use of computational resources. We use one MPI rank per data block and \update{$128$} threads per compute node in all cases. 

\para{Runtime evaluation setup.} 
The existing DHCT construction~\cite{CarrRubelWeber2022a} has three main phases: (1) local contour tree computation, (2) fan-in, and (3) fan-out. For this paper, we added four new phases: (4) augmentation, (5) hypersweep volume computation, (6) branch decomposition, and (7) contour selection and contour extraction, and report runtime for each phase, with any additional costs reported as Other. 

We synchronize all ranks after each phase to facilitate interpretation of the runtime performance. Since these distinct phases depends on completing previous phases, synchronizing all ranks after each phase does not significantly affect performance.  We then compute the maximum runtime of each phase across all ranks, and note that data exchange across ranks is needed for the fan-in phase (2) and all four new phases (4-7) in our implementation.

\begin{figure*}[!t]
\centering
\includegraphics[width=1.0\textwidth]{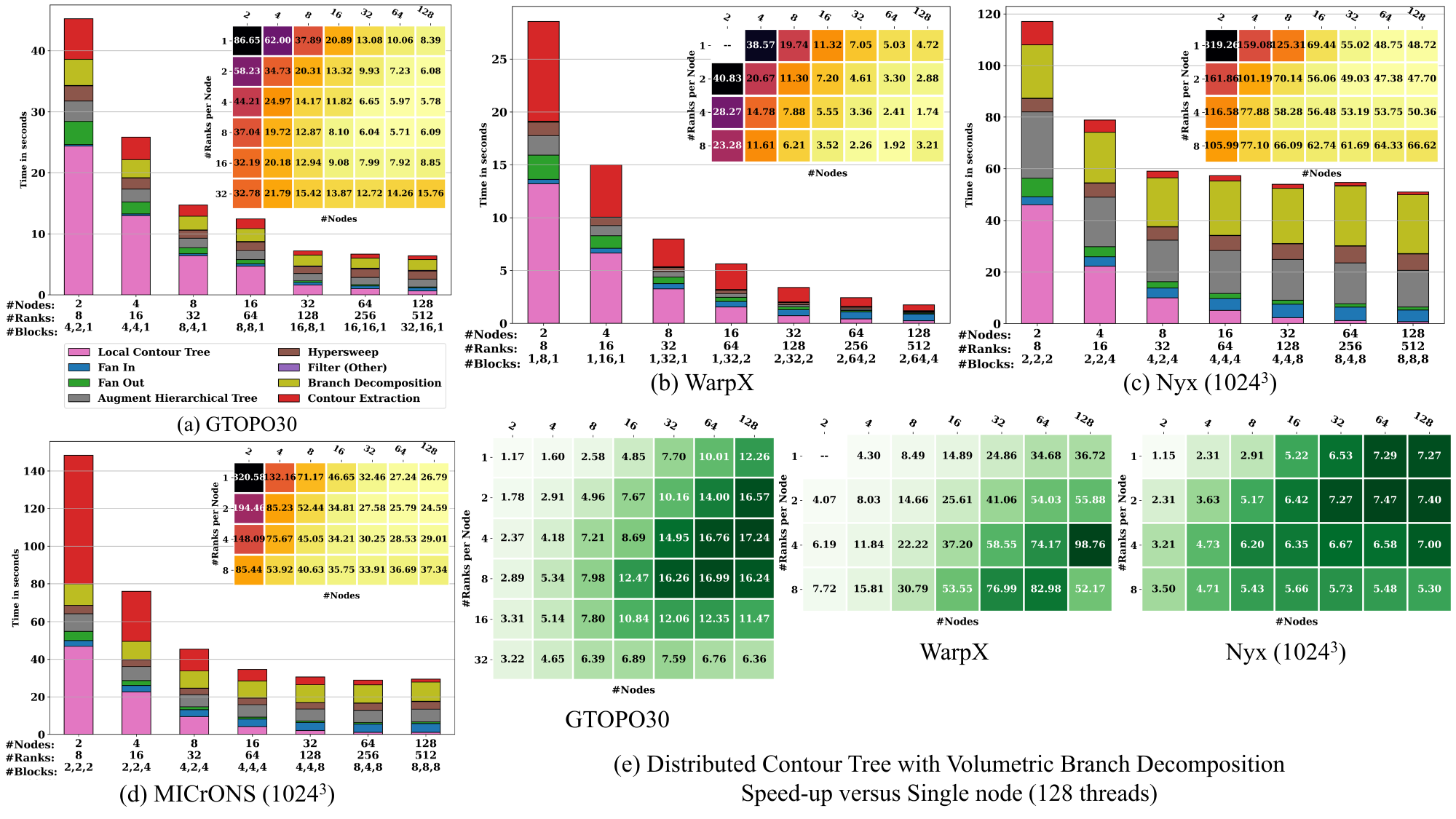}
\vspace{-6mm}
\caption{Strong scaling on Perlmutter using OpenMP. (a)-(d) The bar chart of runtime breakdown by the process for runs with $4$ ranks per node. The heatmaps show the runtime of all node and rank configurations. (e) The speed-ups of the distributed contour tree and branch decomposition, as compared to the single-node PPP implementation.}
\label{fig:perf-general}
\vspace{-4mm}
\end{figure*}

\begin{figure}[!t]
\centering
\includegraphics[width=0.98\columnwidth]{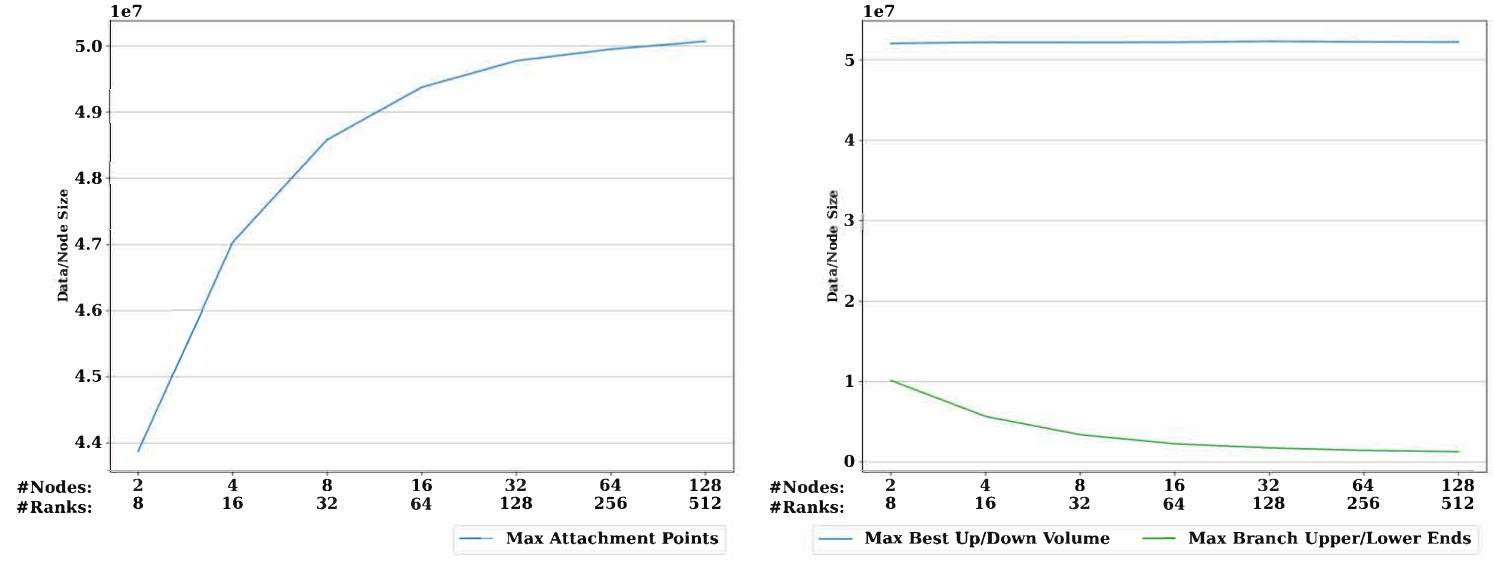}
\vspace{-2mm}
\caption{The maximum size of per-node data exchange for {\Nyx} as the number of nodes increases with $4$ ranks per node. Attachment points are exchanged during augmentation; best up/down volumes and branch upper/lower ends during branch decomposition.}
\vspace{-2mm}
\label{fig:exchange-cost}
\end{figure}

\begin{figure}[!t]
\centering
\includegraphics[width=0.98\columnwidth]{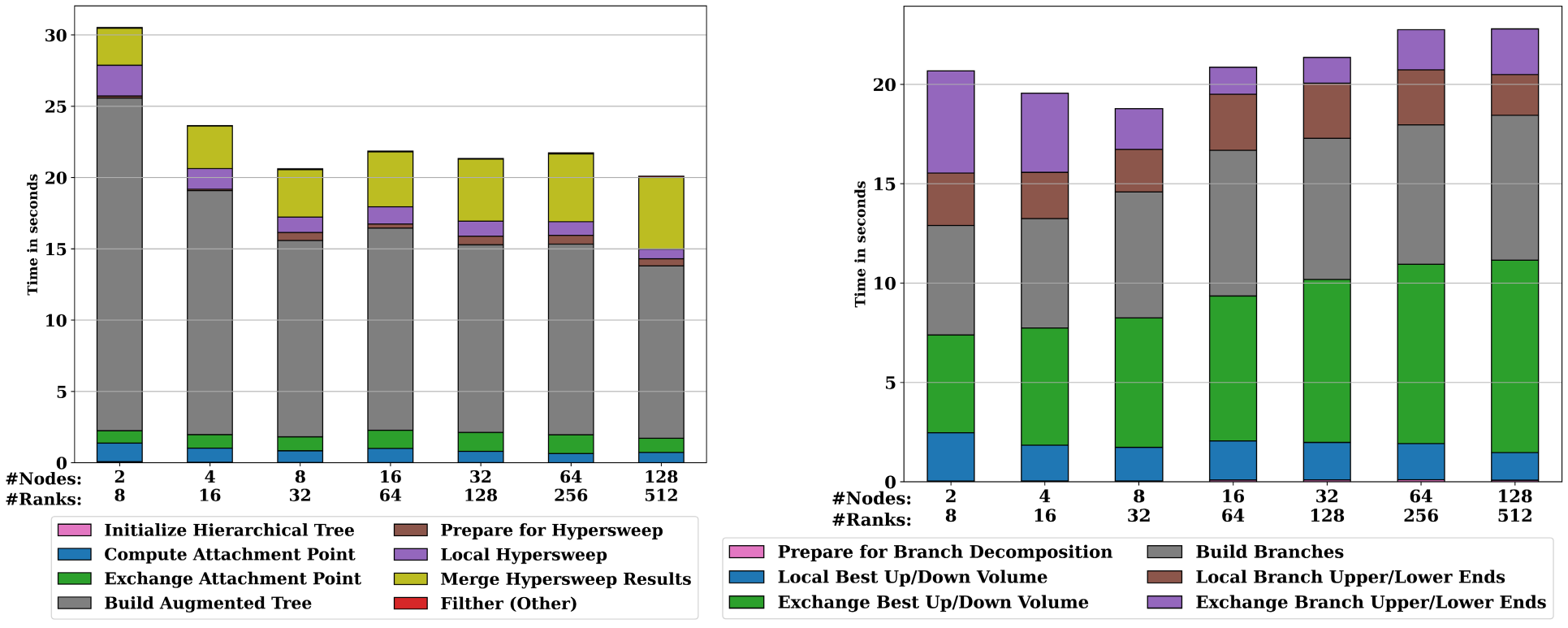}
\vspace{-2mm}
\caption{The stacked runtime bar chart for augmentation and hypersweep (left), and branch decomposition (right) using {\Nyx} at $4$ ranks per node. For each node/rank configuration, the runtime is collected on the rank taking the longest time for the respective phases.}
\label{fig:submodule-evaluation}
\vspace{-6mm}
\end{figure}

\para{Speed-ups.} In~\cref{fig:perf-general} (e) we compare against the single-node PPP~\cite{CarrWeberSewell2021} implementation of the contour tree and volumetric branch decomposition~\cite{HristovWeberCarr2020} in VTK-m. We use OpenMP (\update{128} threads) for single-rank threading. The single-node implementation does not include contour extraction. We, hence, consider for our distributed implementation the total runtime of all phases before contour extraction to compute speed-ups. We do not report speed-ups for the {\MICrONS} dataset, because its topology is too large for single-node computation. 

Overall, our implementation has strong scalability with good speed-ups. Compared to the single-node PPP algorithm with in-node hypersweep implementation, our implementation achieve a maximum speed-up of \update{$\approx 17.24 \times$ for the {\GTO} dataset, $\approx 98.76\times$ for the {\WarpX} dataset, and $\approx 7.47\times$ for the {\Nyx} dataset}. 

\para{Best rank/node configurations.}~Based on physical node configurations, $4$ ranks per node is ideal to utilize the computational resources ($1$ rank per logical CPU). For datasets with complex topology (i.e., {\Nyx} and {\MICrONS}), it is less useful to split data into more ranks due to communication costs, making $2$ ranks per node better. Overall, using $2$ or $4$ ranks per node strikes a balance between utilizing computational resources and reducing communication costs.

\para{Strong scaling performance.}~\cref{fig:perf-general} (a)-(d) shows our performance for all four datasets, where the stacked box plot shows the breakdown of runtime by phase. Contour extraction (red) scales well with the increasing number of nodes. This is because we only need to exchange the information of the selected $k$ branches restricted to $O(rk)$.

While augmentation and hypersweep show improvement in runtime with increasing numbers of nodes for the {\GTO} and {\WarpX} datasets, for the more complex and noisy {\Nyx} and {\MICrONS}, the situation is different: augmentation, hypersweep, and branch decomposition costs remain similar with increasing numbers of nodes. This result follows our analysis in~\cref{sec:complexity}, which shows communication cost staying stable as we subdivide data into smaller blocks, becoming a bottleneck for the scalability.~\cref{fig:exchange-cost} shows the size of exchanged data in multiple phases for {\Nyx}. The maximum workload for a node to exchange attachment points and best down/up volume data here does not decrease with increasing numbers of nodes. 

In contrast, the runtime for the {\WarpX} dataset demonstrates strong scalability. This is probably because the topology for this dataset is relatively clean, leading to a low data exchange workload.

\para{Submodule analysis.} We also analyze scaling behavior for augmentation, hypersweep, and branch decomposition by checking the submodule runtime. Since we do not enforce synchronization after each submodule, we cannot simply take the maximum runtime of submodules across all ranks. Instead, we report the runtime from a single rank with the longest runtime for all submodules in~\cref{fig:submodule-evaluation}. 

Among the first four submodules during augmentation, building the augmented tree takes the longest time. Although this does not involve data exchange, its performance is affected by the number of supernodes in the local tree, which increases as more boundaries are introduced by splitting blocks. From~\cref{fig:submodule-evaluation}, we see its performance benefits from increasing the number of nodes from $2$ to \update{$8$}, but reaching its limit afterward. For the hypersweep phase, the communication cost of merging the hypersweep results becomes the bottleneck. 

In the branch decomposition phase, no submodule demonstrates strong scalability. The local computation for the best up/down volume, building branches, and local branch upper/lower ends are all bounded by the size of known supernodes. Besides, the amount of best up/down volume data to exchange (see~\cref{fig:exchange-cost}) is also not optimized as we increase the number of nodes. 

\para{Highlighted results.} We summarize the performance results:
\begin{itemize}[noitemsep,leftmargin=*]
    \item Our implementation outperforms the state-of-the-art single-node implementation by a large margin, up to \update{$\approx 98.76\times$}.
    \item The scalability of our implementation to extract contours is strong. Contour tree augmentation, hypersweep, and branch decomposition also have strong scalability when the topology of the underlying data is relatively clean. 
    \item On the dataset with complex topology, the boundary information becomes a barrier to further improving the performance of augmentation, hypersweep, and branch decomposition.
\end{itemize}

\section{Conclusion and Discussion}
\label{sec:conclusion}

Our work extends the distributed hierarchical contour tree with augmentation. That is, we insert (regular) attachment points into the contour tree structure to allow geometric measure computation. 
We also develop distributed algorithms to compute geometric measures, branch decomposition, and contour extraction, enabling efficient scientific visualization of contours for large-scale datasets, \update{as well as supporting large-scale contour tree simplification.} Our algorithm demonstrates significant improvement in efficiency over the state-of-the-art. Aside from the high efficiency, our algorithm also enables the distributed storage and computation for large datasets that cannot be handled using the memory of just a single compute node. 

\para{Limitations.}
\update{Our work is not without limitations. The algorithmic performance is significantly influenced by the complexity of the topology of the data. When the topology is complex, the exchanged boundary information increases drastically as the number of blocks increases, limiting the scalability of the algorithm.}

\para{Future work.}
\update{In the future, we intend to scale our implementation further by pre-simplifying the contour tree to reduce the number of attachment points needed in augmentation. 
We also intend to explore other metrics besides volume to enhance the utility of the distributed augmented contour tree.
The appropriate choice of metrics will depend on the scientific applications. 
Moving forward, we expect to apply the distributed hierarchical contour tree and its branch decompositions in topological data analysis and visualization at scale for symmetry detection, feature tracking, and interactive exploration. 
}

\acknowledgments{
This research is supported by the U.S. Department of Energy (DOE), Office of Science, Advanced Scientific Computing Research (ASCR) program and the Exascale Computing Project (17-SC-20-SC), a collaborative effort of the DOE Office of Science and the National Nuclear Security Administration under Contract No.~DE-AC02-05CH11231 to the Lawrence Berkeley National Laboratory.
This research used resources of the National Energy Research Scientific Computing Center (NERSC), a Department of Energy Office of Science User Facility using NERSC award ASCR-ERCAP0026937.
Additionally, Mingzhe Li and Bei Wang are partially supported by DOE DE-SC0021015, National Science Foundation (NSF) IIS-2145499 and NSF IIS-1910733. Hamish Carr is supported by University of Leeds.
}


\bibliographystyle{abbrv-doi-hyperref}
\bibliography{refs-bd.bib}

\clearpage
\newpage
\appendix
\section{\update{Parameter Sensitivity Analysis}}
\label{sec:parameter-sensitivity}

We provide sensitivity analysis for the parameter $k$, the number of branches to select by volume and to extract contours from. 
In particular, we briefly reiterate the asymptotic complexity of the branch extraction phase, followed by experiments on the {\WarpX} and {\Nyx} datasets with a spectrum of values of $k$. 

\subsection{Algorithmic Complexity} 
\label{sec:parameter-sensitivity-complexity}

We have discussed the complexity of algorithmic steps during the contour extraction phase in \cref{sec:complexity}, in which we use $N$ as the number of mesh (regular) vertices, $t$ the size of the contour tree, $B$ the number of branches in the branch decomposition, and $r$ the number of rounds of distributed communication.
Among these steps, identifying the top $k$ branches by volume takes $O(k)$ work in $O(1)$ time, and computing the branch relations among them takes $O(t \lg t + k \lg t)$ work in $O(\lg t)$ time.  Since we assume that $k \ll t$, we expect that $k$ will not affect this stage significantly.

In order to extract a representative contour from each branch, we need to iterate over $O(k)$ isovalues (see~\cref{fig:isosurface-example}) and compute their contours.  We start by looking up the superarc on each branch which the contour maps to. 
This step costs $O (k \lg{t})$ time in our serial implementation, which could be reduced to $O (\lg{t})$ time with $O(k \lg{t})$ work in parallel.

Thereafter, the actual contour extraction relies on visiting every cell in the mesh once for each desired contour, determining whether the isosurface at that value passes through the cell and, if so, testing whether it maps to the right contour.  At present, this step costs $O(N \lg{t})$ work and $O(\lg{t})$ time in parallel, with the log factor deriving from the search for the superarc.  We therefore expect this step to be the principle bottleneck as $k$ increases.

The communication cost during contour extraction is $O(rk)$, where $k \leq B$. This cost is much less than the communication cost during augmentation and is unlikely to be the performance bottleneck.

\subsection{Experimental Configuration}
We fix the experimental configuration to be $16$ nodes with $4$ ranks per node for both {\WarpX} and {\Nyx} datasets. We collect the runtime and workload using $k 
\in \{10^1, 10^2, 10^3, 10^4\}$. 

We collect the runtime of the following steps: 
\begin{enumerate}[noitemsep]
    \item[1.] Locally selecting the top $k$ branches by volume;
    \item[2.] Exchanging the branches between blocks;
    \item[3.] Computing the branch decomposition tree;
    \item[4.] Extracting representative contours.  
\end{enumerate} 
Among the above four steps, steps 2 and 3 include communication across blocks, whereas steps 1 and 4 are local operations (within blocks).

Since the contour extraction is expected to be the dominant cost, we collect the number of cells extracted both before and after testing for whether they map to the desired superarcs.

\subsection{Performance Results}
\label{sec:parameter-sensitivity-performance}

\para{Runtime.} 
As predicted in~\cref{sec:parameter-sensitivity-complexity}, increasing the value of $k$ does not have a huge impact on the performance of computing the top $k$ branches by volume or the branch decomposition tree (steps 1 and 3). 
 
\begin{figure}[!ht]
\centering
\includegraphics[width=\columnwidth]{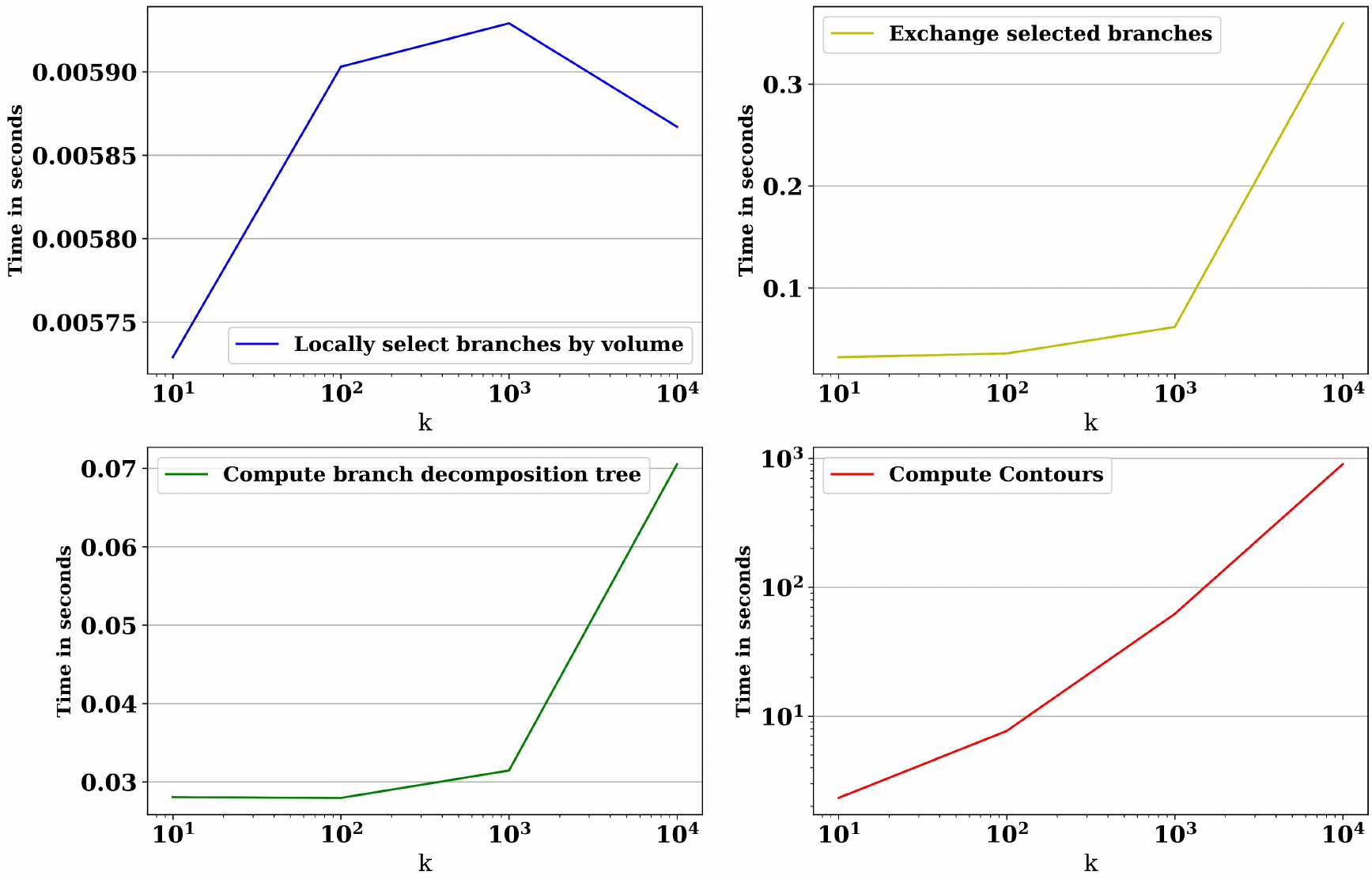}
\vspace{-6mm}
\caption{Runtime of {\WarpX} dataset for steps in the contour extraction phase with $k$ ranging from $10^1$ to $10^4$. }
\label{fig:sensitivity-warpx-runtime}
\vspace{-2mm}
\end{figure}

As shown in~\cref{fig:sensitivity-warpx-runtime}, for the {\WarpX} dataset, the runtime of steps 1 and 3 remain comparable as $k$ varies from $10$ to $10^4$. The runtime of step 2 is within $0.4$ second when $k$ reaches $10^4$. 
On the other hand, the runtime to extract representive contours for the top $k$ branches (step 4) grows linearly as $k$ increases, and reaches $902$ seconds when $k=10^4$. 
Thus, step 4 becomes the dominant factor across all steps.

\begin{figure}[!ht]
\centering
\includegraphics[width=\columnwidth]{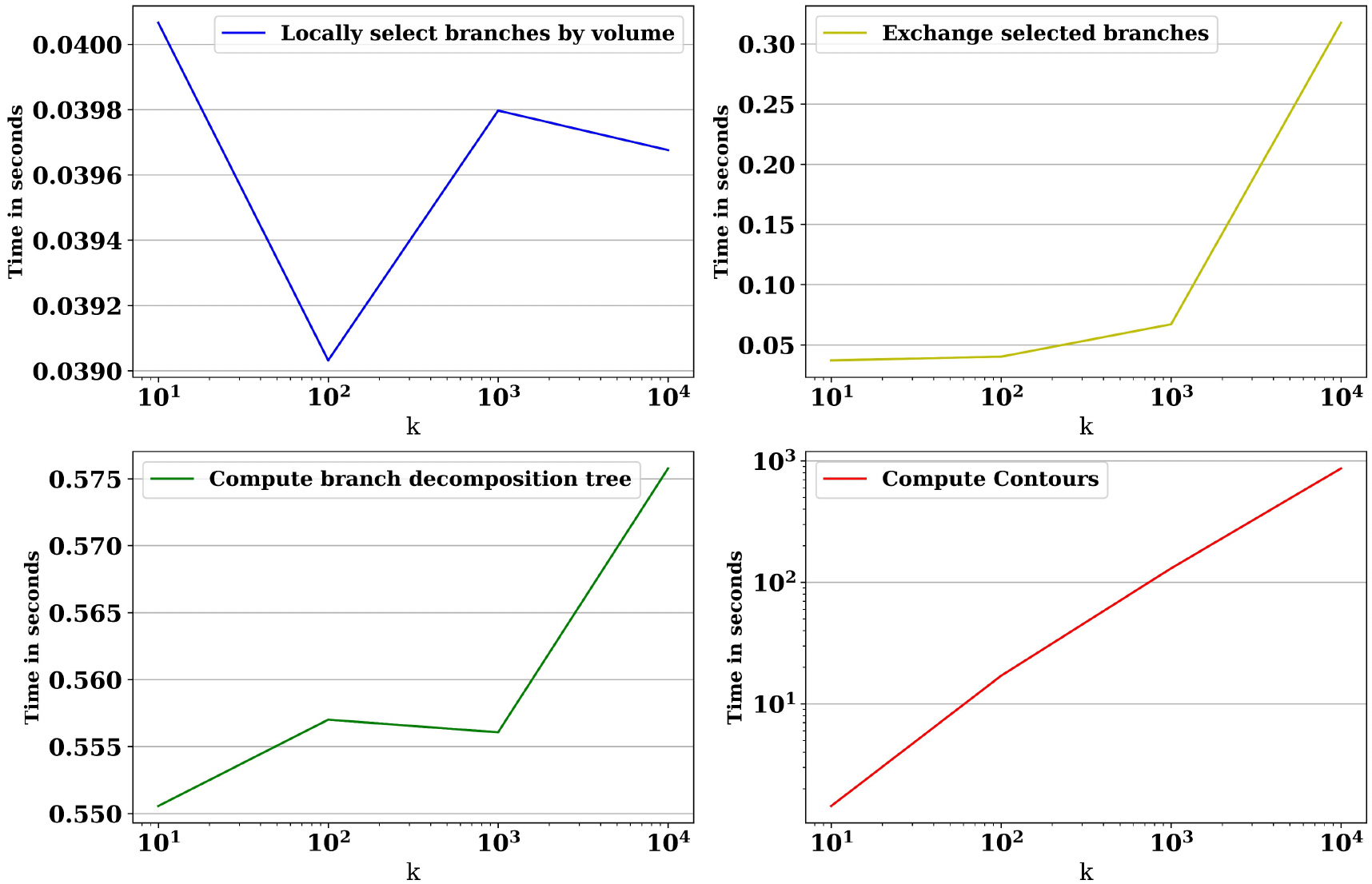}
\vspace{-6mm}
\caption{Runtime of {\Nyx} dataset for steps in the contour extraction phase with $k$ ranging from $10^1$ to $10^4$. }
\label{fig:sensitivity-nyx-runtime}
\vspace{-2mm}
\end{figure}

The runtime performance for the {\Nyx} dataset (see~\cref{fig:sensitivity-nyx-runtime}) is similar: as $k$ increases above $10$, step 4 becomes the dominant factor in performance. 
What is slightly different is that the runtime of steps 1 and 3 for the {\Nyx} dataset is higher than the {\WarpX} dataset, due to its higher topological complexity. 

\begin{figure}[!ht]
\centering
\includegraphics[width=\columnwidth]{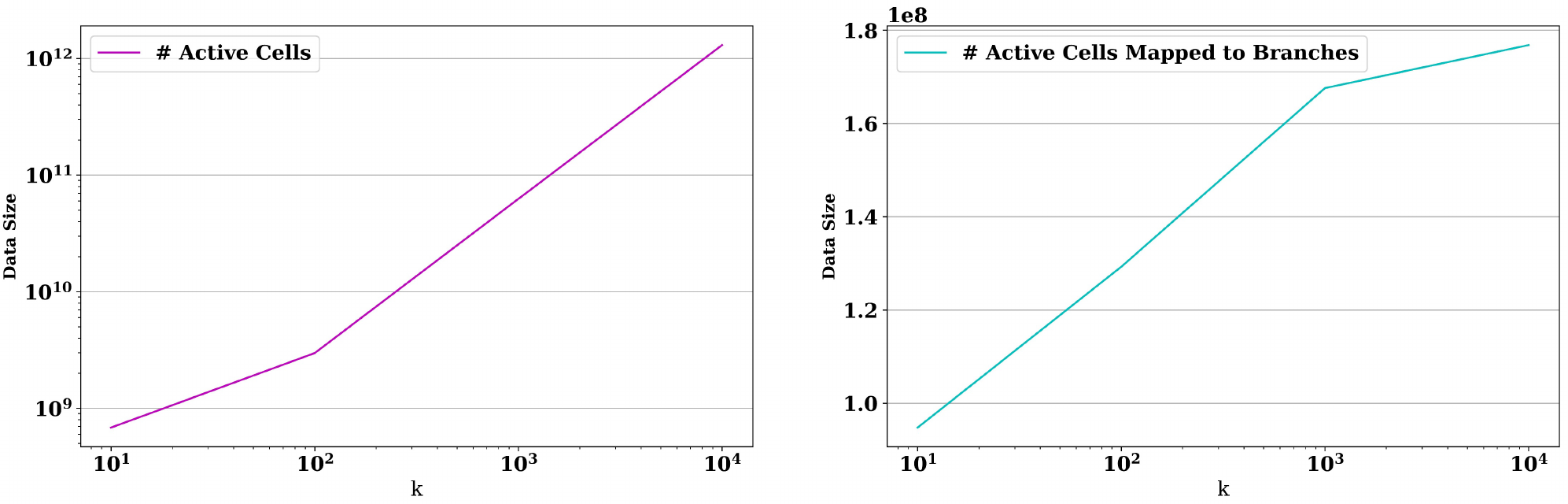}
\vspace{-4mm}
\caption{Workload of {\WarpX} dataset for extracting contours. Left: the number of active cells from the contours. Right: the number of active cells mapped to the selected branches.}
\label{fig:sensitivity-warpx-stats}
\end{figure}

\begin{figure}[!ht]
\centering
\includegraphics[width=\columnwidth]{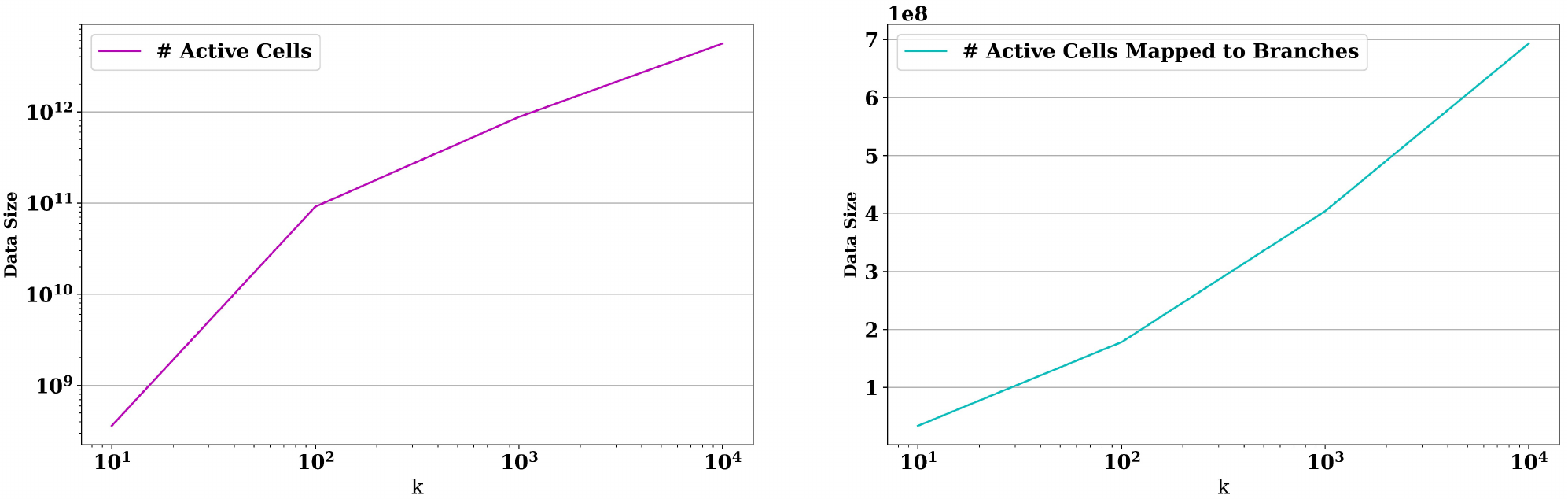}
\vspace{-4mm}
\caption{Workload of {\Nyx} dataset for extracting contours. Left: the number of active cells from the contours. Right: the number of active cells mapped to the selected branches.}
\label{fig:sensitivity-nyx-stats}
\end{figure}

\para{Workload.} 
We diagnose the performance bottleneck of step 4 by checking its workload. For both {\WarpX} and {\Nyx} datasets, the number of active cells grows linearly in $k$; see~\cref{fig:sensitivity-warpx-stats} and~\cref{fig:sensitivity-nyx-stats}. 
This is because we need to compute the superarcs that the active cells map to for each isovalue.

However, the growth of active cells mapping to the selected branches is much slower. 
For example, for the {\WarpX} dataset, there are $9.47\times 10^7$ active cells mapping to the top $11$ branches ($k=10$, adding the main branch). When $k=10^4$, there are $1.77\times 10^8$ active cells mapping to the $10001$ branches; less than half of them map to the $9990$ newly added branches. 
This result implies inefficiency in the computation: for small-volume branches, the surface to be extracted may have only a few triangles, at which point it would be more efficient to use other methods of contour extraction, an optimization that we have not yet implemented.

\subsection{Discussion on Parameter Choices}
\para{Choosing the parameter.} Our primary reason to select a small value of $k$ was to produce visualizations illustrating the algorithms. 
As~\cref{sec:parameter-sensitivity-performance} suggests, contours mapping to small-volume branches are usually small in sizes, making them hard to see.  In addition, since we use volume as the metric for branch decomposition, these contours from small-volume branches may be of less interest. 
Moreover, when $k$ becomes too large, it may be challenging for users to identify contours of interest in the visualization. 

\para{Potential optimization.} 
One idea to optimize the performance of step 4 is to reduce the number of iterations by aggregating the isovalues and removing duplicates to avoid repetitive computations of active cells with the same isovalue, which we leave for future work.


\end{document}